\documentclass[twocolumn,showpacs,preprintnumbers,amsmath,amssymb]{revtex4}
\usepackage{graphicx}
\usepackage{dcolumn}
\usepackage{bm}

\def\R{\partial}
\def\th{\theta}

\def\ep{\varepsilon}
\def\ph{\varphi}
\def\Ps{\varPsi}

\begin{document}
\title{Orienting coupled quantum rotors by ultrashort laser pulses}
\author{Hiroyuki Shima and Tsuneyoshi Nakayama}
\affiliation{Department of Applied Physics, Graduate School of Engineering,
Hokkaido University, Sapporo 060-8628, Japan}
\date{\today}

%%%%%%%%%%%%%%%%%%%%%%%%%%%%%%%%%%%%%%%%%%%%
%
%  Abstract
%
%%%%%%%%%%%%%%%%%%%%%%%%%%%%%%%%%%%%%%%%%%%%

\begin{abstract}

We point out that the non-adiabatic orientation of quantum rotors,
produced by ultrashort laser pulses,
is remarkably enhanced by introducing dipolar interaction between the rotors.
This enhanced orientation of quantum rotors is 
in contrast with the behavior of classical paired rotors,
in which dipolar interactions prevent the orientation of the rotors.
We demonstrate also that a specially designed sequence of pulses
can most efficiently enhances the orientation of 
quantum paired rotors.

\end{abstract} 

\pacs{32.80.Lg, 02.30.Yy, 33.80.Rv, 42.50.Vk, 42.50.Ct}

%\keywords{ }

\maketitle

\section{Introduction}

Considerable attention has been paid to the ability of
intense laser fields to align or orient polar molecules;
in such fields, the molecules experience torque arising
from the dipolar interaction with electric fields \cite{RMP}.
One approach to achieve adiabatic molecular alignment 
is to use a nanosecond-pulse laser \cite{Ortigoso, Larsen,Friedrich}.
While adiabatic alignment disappears after the pulse is turned off, 
ultrashort pulses (100 fs or less) can excite rotational wave packets of quantum rotors,
thus yielding a noticeably aligned shape after the pulse is off
\cite{Heritage,Bandrauk,Felker,Seideman2,Cai,Rosca,Litvinyuk}.
Furthermore, a specially designed sequence of pulses 
is known to achieve an enhanced angular focusing 
in quantum rotors \cite{Averbukh_2001,Averbukh_2003};
this has been realized experimentally in optical lattices \cite{Oskay}.
These types of alignment, which differ from the adiabatic,
are important for manifold applications 
requiring transient molecular alignment under field-free conditions,
such as the generation of laser pulses \cite{Bartels,Kalosha} 
and the control of high harmonic generation as a source of 
coherent radiation \cite{Velotta}.

In most studies of the molecular alignment,
%molecules driven by short laser pulses have been assumed to be
%independent of each other,
%interaction between molecules has not been taken into account so far.
dipolar interaction between polar molecules has been neglected.
In this case,
the rotational dynamics of polar molecules can be simply analyzed 
by using an isolated kicked-rotor model \cite{Haake}. 
The quantum kicked rotor and its classical analog have long served as a paradigm 
for quantum and classical chaos.
In contrast, the quantum dynamics of {\it interacting} kicked rotors 
has been less well studied so far, though intriguing phenomena
of interacting quantum rotors are expected to emerge.
Very recently, a study of the center-of-mass motion in two coupled kicked rotors
has revealed that the decoherence effect induced by the internal degree of freedom
enhances the quantum-classical correspondence in the dynamics of rotors \cite{Park}.
In addition, anomalous dielectric responses 
have been pointed out in two coupled dipolar rotors \cite{Shima}.
These results imply that coupled kicked rotors, 
even in {\it only two} rotors, exhibit peculiar behaviors
different from those of isolated kicked rotors.

In the present work, we theoretically investigate the quantum dynamics 
of coupled (paired) kicked rotors subject to an ultrashort laser pulse 
($\delta$-function kick).
We find that the dipolar interaction between rotors
remarkably enhances the transient orientation of paired rotors 
produced by a $\delta$-kick.
This enhancement of the orientation of {\it quantum} rotors
contrasts with the results 
with the {\it classical} treatment for paired rotors;
in the latter case, the dipolar interaction
inevitably hinders the orientation of coupled rotors.
Furthermore, we demonstrate that the orientation of paired rotors 
can be further enhanced by applying the accumulative squeezing scheme 
proposed in Ref.~\cite{Averbukh_2001}.
Our findings enlighten the study of transient orientation of 
interacting polar molecules.

This paper is organized as follows.
Section II describes the Hamiltonian of the paired-rotor system 
together with its analytical solutions for eigenenergies and their eigenfunctions.
The time development of paired rotors for {\it post}-kicked times is given in this section.
Section III analyzes the time dependence of the orientation factor
in both quantum and classical paired rotors.
The physical origin of the enhanced orientation in quantum rotors 
is discussed on the basis of the spatial profile of the probability density 
of paired-rotor wavefunctions.
Section IV gives the conclusion.
The paper contains two Appendices with the details of the calculations.

%==================================================

\section{Paired-rotor systems}

\subsection{The Hamiltonian}

Suppose that two rotors carrying dipole moments $\bm{\mu}$ are 
arranged as shown in Fig. 1.
For arrangement (a), both rotors rotate in a plane, 
while in the case of (b), both rotors belong to an identical rotation axis \cite{experiment}.
The Hamiltonian for the system is given by
$H = H_1 + H_2+ W_{12}$, where $H_i$ is the Hamiltonian 
for the $i$-th kicked rotor and $W_{12}$ represents dipolar interaction between rotors.
The term $H_i$ can be written as
\begin{equation}
H_i = \frac{{L_i}^2}{2I} + V(\th_i,t),
\label{eq01}
\end{equation}
where $L_i$ is the angular momentum operator and $I$ is the moment of inertia of the rotor.
When rotors are driven by a linearly polarized field, we can set
\begin{equation}
V(\th_i,t)=-\mu E(t)\cos(\th_i),
\label{eq02}
\end{equation}
%
%------------------------------
\begin{figure}[ttt]
\hspace*{-0.5cm}
\includegraphics[width=10cm]{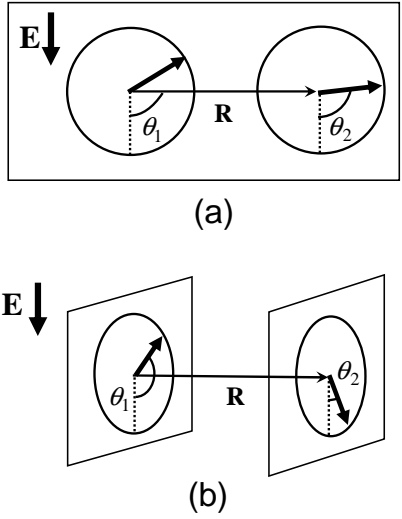}
\caption{
Definitions of arrangements of two dipolar rotors:
(a) Two dipoles rotate in an identical plane. 
(b) Two rotors belong to an identical rotation axis.
In both cases, the rotors are separated by $R$
and interact via the dipolar interaction $W_{12}$.
The direction of the electric field $\bm{E}(t)$ is defined as shown.}
\label{fig1}
\end{figure}
%--------------------
%
where $E(t)$ is the field amplitude of short laser pulses.
The direction of the field is fixed as shown in Fig.~1,
causing the angular focusing of paired rotors at $\th_1=\th_2=0$.
Assuming that the rotational radius of rotors is sufficiently 
smaller than the separation $R$ between rotors, 
the dipolar interaction $W_{12}$ is expressed by
\begin{equation}
W_{12} = \frac{1}{4\pi\ep R^3}
\left[ \bm{\mu}_1\cdot \bm{\mu}_2 
      -3\frac{(\bm{\mu}_1\cdot \bm{R})(\bm{\mu}_2\cdot \bm{R})}
      {R^2}\right]
\label{eq03}
\end{equation}
with the dielectric constant $\ep$.
The vector $\bm{R}$ connects two rotational centers as denoted in Fig.~1.
For simplicity, we rewrite the term (\ref{eq03}) as 
\begin{equation}
W_{12}=E_D\cdot F(\th_1,\th_2)
\label{eq04}
\end{equation}
with the definition,
\begin{equation}
E_D=\frac{\mu^2}{4\pi\ep R^3}.
\label{eq05}
\end{equation}
The quantity $E_D$ determines the magnitude of the dipolar interaction 
between rotors, and plays a key role in the dynamics of coupled rotors,
as we discuss later.
The explicit form of $F(\th_1,\th_2)$ is obtained straightforwardly from Fig.~1.
The arrangement (a) gives the form,
\begin{equation}
F(\th_1,\th_2) = \cos\th_1 \cos\th_2 - 2 \sin\th_1 \sin\th_2,
\label{eq06}
\end{equation}
while, for arrangement (b), we have
\begin{eqnarray}
F(\th_1,\th_2) &=& \cos\th_1 \cos\th_2 + \sin\th_1 \sin\th_2 \nonumber \\
 &=& \cos(\th_1-\th_2).
\label{eq07}
\end{eqnarray}

%============================================

\subsection{Eigenenergies and their eigenfunctions}

In the absence of the field $E(t)$, eigenstates of paired rotors
are analytically obtained by transforming variables into
$\xi=(\th_1+\th_2)/2$ and $\eta=(\th_1-\th_2)/2$.
Substituting them into Eqs. (\ref{eq01}) and (\ref{eq04}), 
the Hamiltonian $H$ is separated as \cite{Shima},
\begin{equation}
H = H_{\xi}+H_{\eta},
\label{eq08d}
\end{equation}
\begin{equation}
H_{\alpha} = -\frac{E_K}{2} \frac{\R^2}{\R \alpha^2}
+ E_D c_{\alpha} \cos 2(\alpha+\alpha_0); \;\; \alpha=\xi,\eta.
\label{eq08}
\end{equation}
Here the quantity $E_K \equiv \hbar^2/(2I)$ represents the kinetic energy.
The parameters $(c_{\xi},c_{\eta},\xi_0,\eta_0)$ equal
$(3/2, 1/2, 0, \pi/2)$ for arrangement (a) and $(0,1,0,0)$ for (b).
The separability of the Hamiltonian $H$ allows us 
to write the paired-rotor wavefunction in the form
$\Ps(\xi,\eta) = \ph_{\xi}(\xi)\ph_{\eta}(\eta)$.
Consequently, the Schr\"odinger equation of paired rotors
$H\Ps(\xi,\eta)=E\Ps(\xi,\eta)$ can be decomposed into 
two independent eigenvalue equations expressed by
\begin{equation}
\frac{\R^2\ph_{\alpha}}{\R \alpha^2} +
\left[\ep_{\alpha} - 2 v_{\alpha} \cos 2(\alpha+\alpha_0) \right]\ph_{\alpha} = 0;
\;\; \alpha=\xi,\eta,
\label{eq09}
\end{equation}
where $\ep_{\alpha}=2E_{\alpha}/E_K$ and $v_{\alpha}=c_{\alpha} E_D/E_K$.
The solution of Eq.(\ref{eq09}) is given by
the Mathieu function \cite{Whittaker},
whose explicit forms are given in Appendix A.
The eigenenergies $E$ of paired rotors are thus expressed by
$E=(\ep_x+\ep_y)E_K/2$.

%============================================

\subsection{Time development of wavefunctions}

Let us consider the time development of the wavefunction
in paired-rotor systems
after a $\delta$-function kick at $t=\tau$.
The wavefunction $\Ps(\th_1,\th_2,\tau^{+})$ immediately
after the kick is related to that just before the kick,
$\Ps(\th_1,\th_2,\tau^{-})$, with a phase determined
by
\begin{eqnarray}
& &\Ps(\th_1,\th_2,\tau^{+}) \nonumber \\
& &= \exp\left[ \int_{-\infty}^{\infty} -\frac{i}{\hbar}
\left\{ V(\th_1,t)+V(\th_2,t) \right\} dt \right] \Ps(\th_1,\th_2,\tau^{-}).
\label{eq10}
\end{eqnarray}
Substituting the definition (\ref{eq02}) into Eq.~(\ref{eq10}) 
and transforming variables $(\th_1,\th_2) \to (\xi,\eta)$,
we obtain
\begin{equation}
\Ps(\xi,\eta,\tau^{+}) =
\sum_{n=-\infty}^{\infty} i^n J_n \left( \frac{2P}{\hbar}\cos \xi \right)
\exp(-in\eta) \Ps(\xi,\eta,\tau^{-}),
\label{eq11}
\end{equation}
where $J_n(z)$ is the Bessel function of $n$-th order.
The quantity,
\begin{equation}
P=\int_{-\infty}^{\infty} \mu E(t) dt,
\label{eq12}
\end{equation}
represents the strength of the pulse.
In actual calculations for Eq.~(\ref{eq11}),
the summation of $n$ can be truncated at the finite value $\pm n_c$,
since the magnitude of the Bessel function 
$J_n \left( \frac{2P}{\hbar}\cos \xi \right)$ rapidly decays
with increasing $|n|$.
The time development of paired rotors for {\it post}-kicked times
is described by
\begin{equation}
\Ps(\xi,\eta,\tau^{+}+t)
= \exp\left[ -\frac{i}{\hbar} (H_{\xi}+H_{\eta})t \right] 
\Ps(\xi,\eta,\tau^{+}).
\label{eq13}
\end{equation}
The right-hand side of Eq.~(\ref{eq13})
can be analytically calculated by expanding
the function $\Ps(\xi,\eta,\tau^{+})$ by the Mathieu function.
The details of the calculation are presented in Appendix B.
In the following, we set $\tau=0$, and take $E_K$ and $\hbar/E_K$
as units of energy and time, respectively.
The initial state just before the kick
is fixed in the ground state.

% ==========================================

\section{Orientation of paired rotors}

\subsection{The orientation factor}

%----------------------------------------------------------------------------------
\begin{figure}[ttt]
\includegraphics[width=8.8cm]{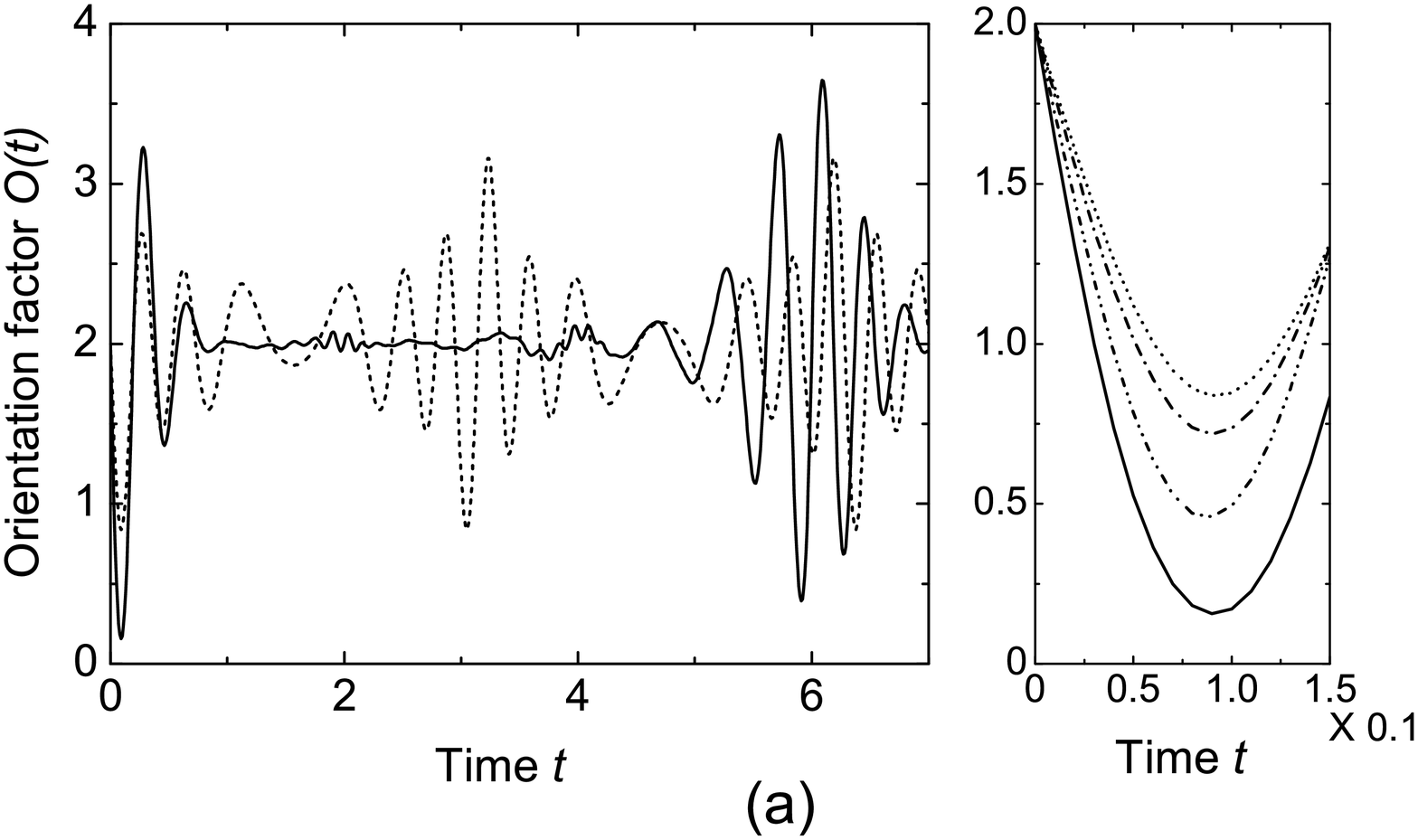}
\includegraphics[width=8.8cm]{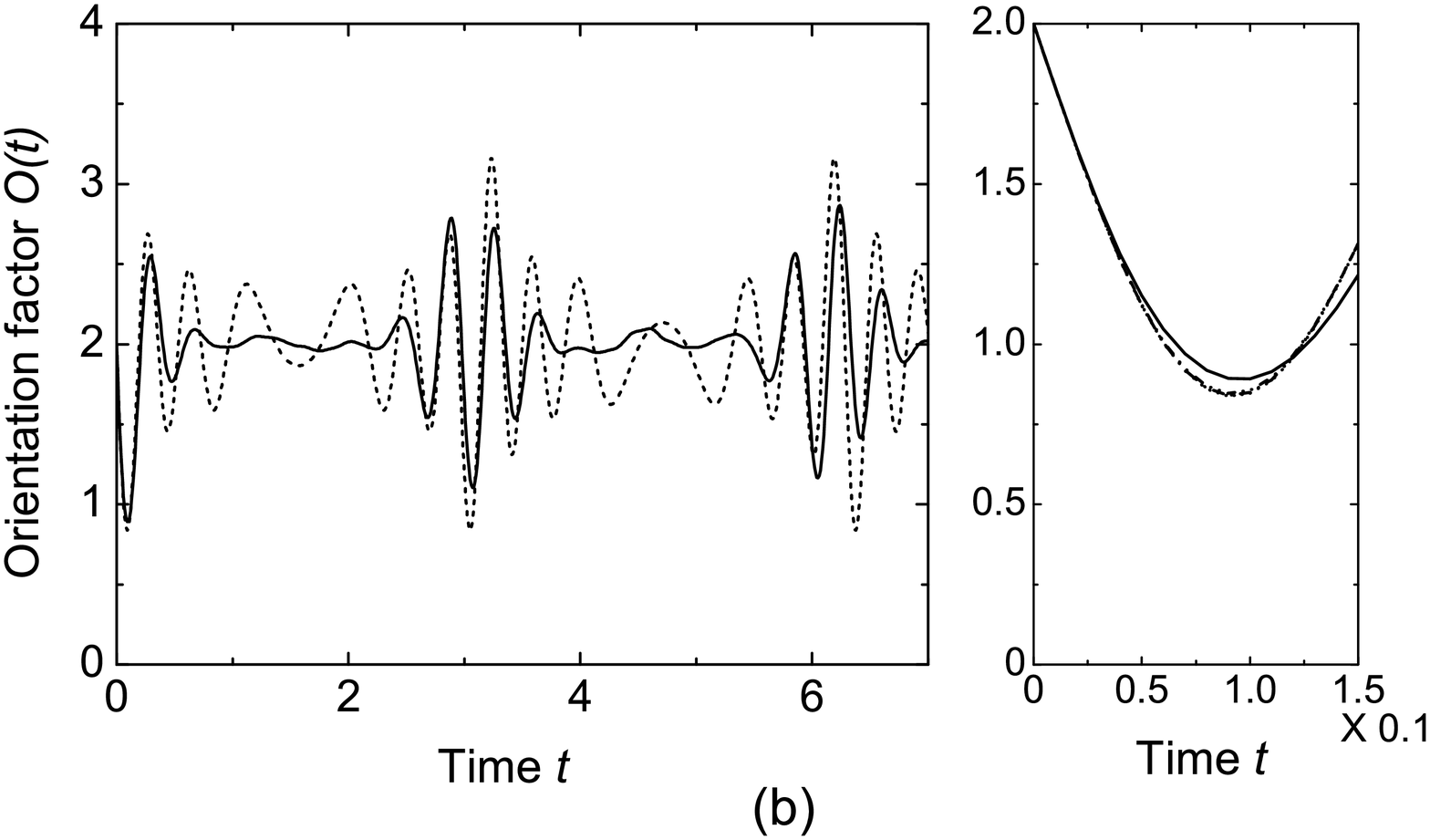}
\caption{The orientation factor $O(t)$ in quantum paired rotors.
The figures (a) and (b) correspond to arrangements (a) and (b) in Fig.~1.
The quantity $\hbar/E_K$ is taken as the unit of time.
Left: The time dependence for $O(t)$ within a long time scale.
The parameter $\Gamma$ is varied as $\Gamma=0$ (dotted line)
and $\Gamma=30.0$ (solid line)
fixing the kick strength $P=10$.
Right: The behavior of $O(t)$ within a short time scale.
$\Gamma$ is varied as $\Gamma=0$ (dotted line),
$\Gamma=1.0$ (dashed-dotted),
$\Gamma=3.0$ (dashed-dotted-dotted),
and $\Gamma=30.0$ (solid).}
\label{fig2}
\end{figure}
%--------------------

The degree of orientation for paired rotors
is characterized by the orientation factor
$O(t)=\langle 2-\cos\th_1-\cos \th_2 \rangle$,
where the angular bracket indicates to take the expectation value.
The factor $O(t)$ tends to be zero when the orientation of the rotors 
in the field direction of $\bm{E}$ is perfect.
On the other hand,
the factor equals $2$ when the amplitude of wavefunction  $|\Ps(\th_1,\th_2)|$
is uniformly distributed
in the $\th_1$-$\th_2$ space.
The strength of the dipolar interaction is characterized by the parameter
$\Gamma=E_D/E_K$, {\it i.e.}, the ratio of the interaction energy $E_D$
to the kinetic energy $E_K$.

Figure 2 (a) and (b) show the time dependence for the orientation factor $O(t)$
for paired rotors fixing the kick strength $P=10$.
Indices (a) and (b) in Fig.~2 correspond to arrangement of rotors
(a) and (b) in Fig.~1.
In both (a) and (b),
the left figure plots the orientation factor for a long time scale, $0\le t \le 7.0$,
while the right one does so for a short time scale, $0\le t \le 0.15$.
We first discuss the two figures on the left,
where two values of $\Gamma$ are taken;
the dotted lines display the orientation factor
for $\Gamma=0$, and the solid lines display that for $\Gamma=30.0$.
For $\Gamma=0$, the time dependence for $O(t)$ for arrangement (a)
is identical to that for (b),
because the two rotors are no longer correlated via dipolar interaction.
In this case, the orientation factor yields 
a simple form \cite{PRA} of
\begin{equation}
O(t)=2-2 J_1\left( 2P \sin t \right)
\label{eq14}
\end{equation}
with the lowest value of $O(t_c)=0.836$ 
at the focal time $t=t_c=9.2 \times 10^{-2}$.

For finite $\Gamma$'s,
the orientation factor exhibits somewhat complicated behavior different from that for
isolated rotors.
The left two figures in Fig.~2 exhibit the difference of
the time dependence for $O(t)$ between the case of $\Gamma=30.0$ and that of $\Gamma=0$.
For arrangement (a), 
the magnitude of $O(t)$ for $\Gamma=30.0$ exceeds that for $\Gamma=0$
at a time $t\approx 0$ and $t\approx 2\pi$.
We must notice
that the lowest value of $O(t)$
for $\Gamma=30.0$ located at $t=t_c\approx 0.1$
is remarkably smaller than that for $\Gamma=0$.
This indicates that, in arrangement (a),
the orientation of paired rotors is enhanced by introducing 
strong dipolar interaction.
For arrangement (b), in contrast,
the magnitude of $O(t)$ for $\Gamma=30.0$ does not exceed
that for $\Gamma=0$ at any $t$.
In addition, the lowest value of $O(t)$ at $t\approx 0.1$ seems to be
invariant to the change of $\Gamma$,
implying that the orientation of paired rotors in arrangement (b)
is not much affected by dipolar interaction.

In order to examine the effect of the interaction for the lowest value of $O(t)$,
we investigate in detail the behavior of $O(t)$ around the focal time $t_c$
with varying $\Gamma$.
The calculated results are shown in the right two images in Fig.~2,
where the value of $\Gamma$ is increased from $\Gamma=0$ (dotted line)
up to $\Gamma=30.0$ (solid line).
In case (a),
the increase in $\Gamma$ monotonously reduces
the lowest value of $O(t)$.
For $\Gamma=30.0$, the factor eventually takes the lowest value $O(t_c)=0.156$
at the focal time $t_c=9.1 \times 10^{-2}$,
which is much smaller than the lowest value of $O(t)$ for $\Gamma=0$.
In case (b), on the other hand,
the time dependence for $O(t)$ hardly changes with varying $\Gamma$.
We thus conclude that, as far as arrangement (a) is concerned,
the orientation of paired rotors can be efficiently enhanced
by taking into account the dipolar interaction between rotors.
This is one of main findings of the present study.
As we see in the next subsection,
the enhanced orientation in {\it quantum} rotors
cannot be interpreted from the {\it classical} dynamics for paired rotors,
indicating that the enhanced orientation stems from a purely quantum effect.

\subsection{Classical paired rotors}

Before proceeding to a further investigation of {\it quantum} paired rotors,
we consider the effect of dipolar interaction
on the orientation for {\it classical} paired rotors.
For classical kicked rotors,
the term $H_i$ defined in (\ref{eq02}) is rewritten as
\begin{equation}
H_i = \frac{I}{2}\dot{\th}_i^2(t) + V(\th_i,t),
\label{eq15}
\end{equation}
while the interaction term $W_{ij}$ is the same as that defined in (\ref{eq04}).
Under field-free conditions,
the equations of motion for arrangement (a) are given by
\begin{eqnarray}
I\ddot{\th}_1 &=&
\frac{E_D}{2}\left[ 3 \sin(\th_1+\th_2) - \sin(\th_1-\th_2) \right]
\label{eq15a}, \\
I\ddot{\th}_2 &=&
\frac{E_D}{2}\left[ 3 \sin(\th_1+\th_2) + \sin(\th_1-\th_2) \right]
\label{eq15b}.
\end{eqnarray}
By transforming the variables into $q_1=\th_1+\th_2$ and $q_2=\th_1-\th_2$,
we obtain the following equations:
\begin{eqnarray}
\ddot{q}_1(t) &=& 6\Gamma_{cl} \sin q_1(t), \label{eq16} \\
\ddot{q}_2(t) &=& -2\Gamma_{cl} \sin q_2(t). \label{eq17}
\end{eqnarray}
Here we define the parameter $\Gamma_{cl}=E_D/(2I)$,
showing the strength of the dipolar interaction between classical rotors.
For arrangement (b), the same procedure yields
\begin{eqnarray}
\ddot{q}_1(t) &=& 0, \label{eq18} \\
\ddot{q}_2(t) &=& 4\Gamma_{cl} \sin q(t). \label{eq19}
\end{eqnarray}
Solutions of Eqs.~(\ref{eq16})-(\ref{eq19})
are expressed by Jacobi's elliptic functions.
The orientation factor for classical paired rotors
is calculated by
$O(t)= 2- 2 \langle\cos q_1(t) \cos q_2(t) \rangle_{cl}$,
where the bracket $\langle \cdots \rangle_{cl}$
means averaging over initial angles $\th_i(t=0)$ \cite{initial}.

%----------------------------------------------------------------------------------
\begin{figure}[ttt]
\includegraphics[width=8.3cm]{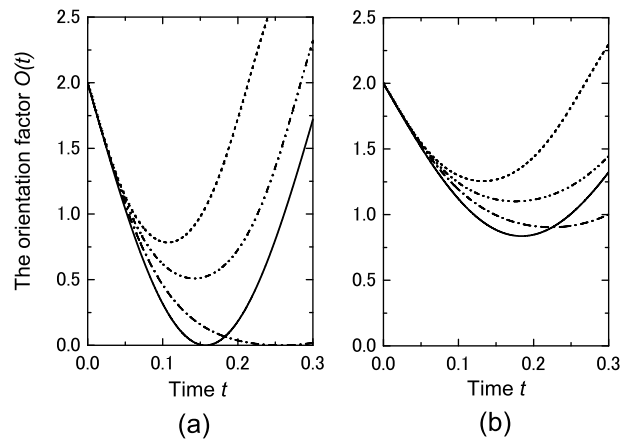}
\caption{The orientation factor $O(t)$ in classical paired rotors.
Figures (a) and (b) correspond to arrangements (a) and (b) in Fig.~1.
The quantity $I/P$ is taken as the unit of time.
The parameter $\Gamma_{cl}=E_D/(2I)$, determining the strength of dipolar interaction,
is varied as $\Gamma_{cl}=0, 15, 30, 45$ from the bottom (solid line) 
to the top (dashed line).}
\label{fig3}
\end{figure}
%--------------------

Figure 3 shows the time dependence of the orientation factor
for the classical paired rotors.
Quantities $I/P$ and $(I/P)^2$ are taken as the unit of time 
and the parameter $\Gamma_{cl}$, respectively.
The value of $\Gamma_{cl}$ is incrementally increased
from $\Gamma_{cl}=0$ (dotted line)
up to $\Gamma_{cl}=45$ (solid line)
as denoted in the figure caption.
We see that the increase in $\Gamma_{cl}$
inevitably raises the minimal value of $O(t)$ in both arrangements (a) and (b).
This leads to the conclusion that, in {\it classical} systems,
the strong dipolar interaction
interferes with the orientation of paired rotors for both arrangements (a) and (b).
The physical interpretation is given as follows.
When two dipolar rotors are assigned in arrangement (a),
strong dipolar interaction forces them to be parallel 
in the direction normal to the field direction (See Fig.~1).
Hence, the interaction hinders the orientation of rotors in the field direction.
For arrangement (b), on the other hand,
two dipolar rotors tend to be anti-parallel to each other.
This increases the minimal orientation factor at the focal time.
As a consequence, the dipolar interaction in {\it classical} systems certainly
prevents the rotors from becoming oriented in the field direction defined in Fig.~1.

These facts naturally lead us to the following question:
Why is it that a strong dipolar interaction can enhance
the orientation of rotors in {\it quantum} systems (See Fig.~2 (a)) ?
Comparing the behavior of $O(t)$ shown in Fig.~3 (a)
with that shown in the right of Fig.~2 (a),
we clearly see that the effect of dipolar interaction on the orientation
of paired rotors differs completely between classical and quantum systems.
To settle this point,
we consider the time development of wavefunctions in the paired-rotor system
as follows.

\subsection{Time development of wavefunctions for post-kicked time}

To understand the mechanism underlying the enhanced orientation
in quantum paired rotors,
we examine the time development of the probability density $|\Ps(\th_1,\th_2,t)|^2$
in the $\th_1$-$\th_2$ space.
Figure 4 (i) and (ii) give contour plots of the probability density
for arrangement (a) with $\Gamma=30.0$.
At $t=0$ (Fig.~4 (i)), the amplitude of $|\Ps(\th_1,\th_2,t)|^2$ is spatially
confined around the two symmetric positions 
$(\th_1,\th_2)=(\pi/2,\pi/2)$ and $(-\pi/2,-\pi/2)$.
When the kick is applied, these two wave packets move toward the origin
$\th_1=\th_2=0$ while retaining their shapes,
and finally collide head-on with each other 
at the origin at a focal time $t=t_c$.
The resultant angular focusing is demonstrated in Fig.~4 (ii),
where the probability density of $|\Ps(\th_1,\th_2)|^2$
is well localized at around the origin.
This transient angular focusing at the origin
enhances the orientation of quantum paired rotors.

%----------------------------------------------------------------------------------
\begin{figure}[ttt]
\includegraphics[width=8.5cm]{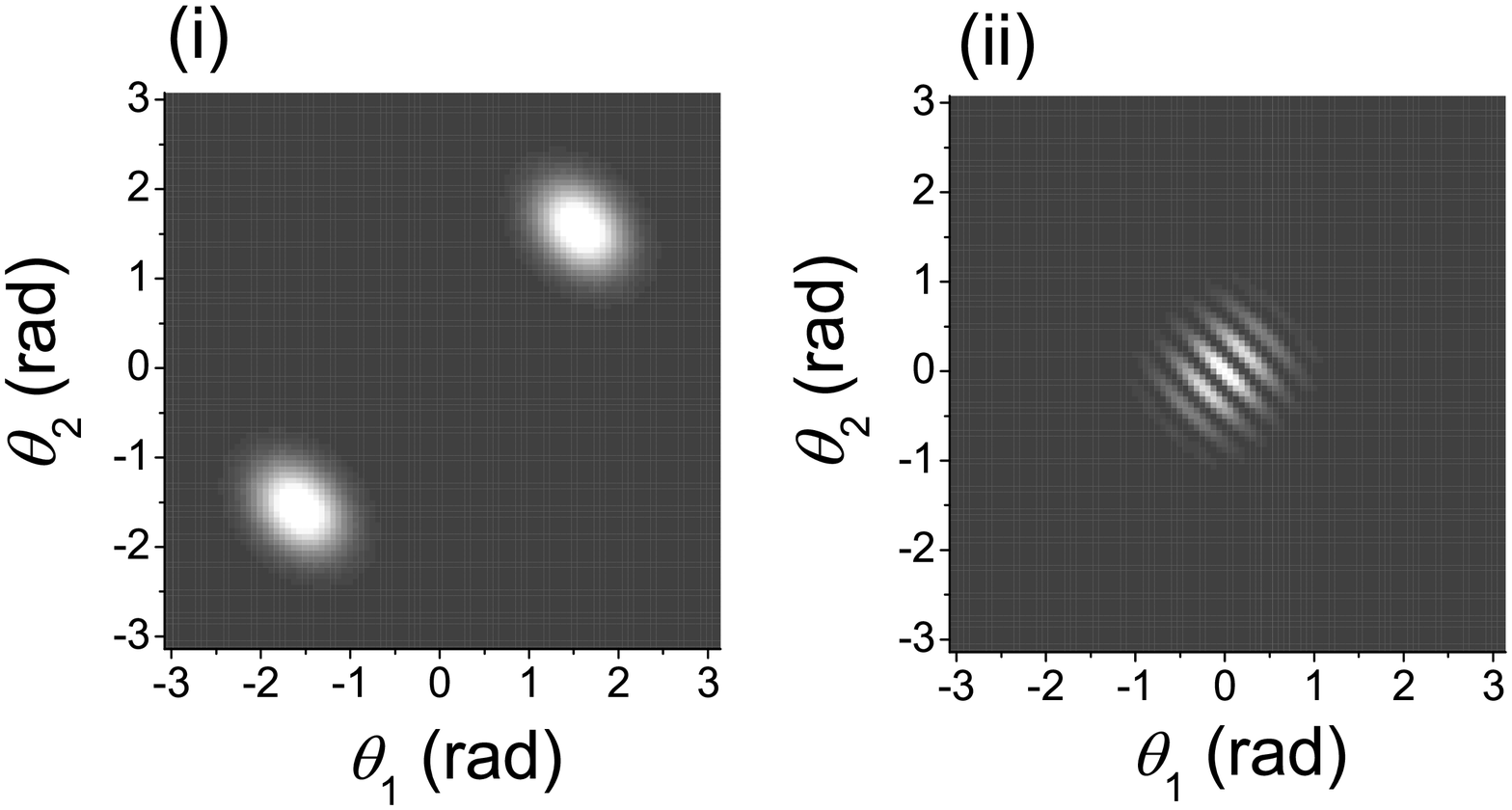}
\includegraphics[width=8.5cm]{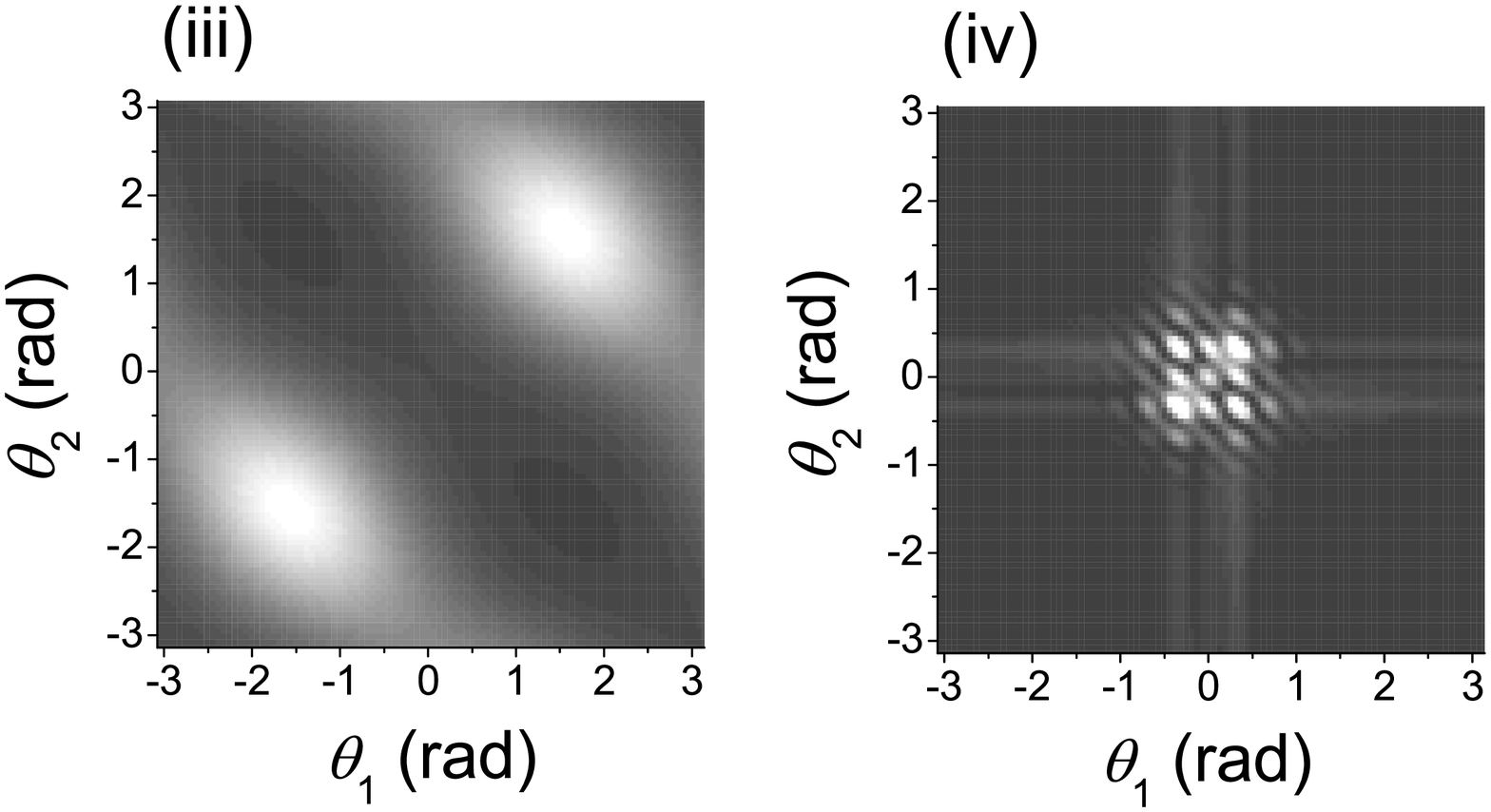}
\caption{Contour plots of the probability density
$|\Ps(\th_1,\th_2,t)|^2$ for arrangement (a) in Fig.~1.
Top: The parameter $\Gamma=30.0$ with
(i) $t=0$, and (ii) $t=t_c$.
Bottom: $\Gamma=1.0$ with (iii) $t=0$, and (iv) $t=t_c$.
White regions indicate large amplitudes of the probability density.}
\label{fig4}
\end{figure}
%--------------------

We must note that angular focusing is enhanced
only when $\Gamma \gg 1$,
namely, when the two rotors are strongly correlated via dipolar interaction.
If the dipolar interaction is sufficiently weak ($\Gamma \le 1$),
the probability density for the initial state
is broadly distributed in the $\th_1$-$\th_2$ plane (Fig.~4 (iii)).
After a kick is applied,
the amplitude of the wavefunction spreads out over the $\th_1$-$\th_2$ space,
followed by the formation of a ``rainbow structure" \cite{Averbukh_2001}
at a focal time $t_c$ (Fig.~4 (iv)).
The degree of angular focusing for the rainbow structure
is obviously inferior to that for the case of strongly interacting rotors
(Fig.~4 (ii)).
In summary, two factors are essential for enhancing the orientation of
paired rotors:
i) The initial state before the kick consists of two wave packets
strongly confined at symmetric positions with respect to the origin,
ii) These wave packets move translationally toward the origin after the kick.
It should be mentioned that the translational motion of two wave packets
after a $\delta$-pulse is not trivial.
We can analytically trace the motion of those wave packets
by calculating the expansion coefficients
$D_{ll'}$ appearing in Appendix B.
Details of the calculations will be published elsewhere \cite{future}.

The spatial profile of the initial eigenstate
is determined by the potential term $W_{12}$.
For arrangement (a), the potential $W_{12}$ as a function of
$\th_1$ and $\th_2$ gives two potential minima, at 
$(\th_1,\th_2)=(\pi/2,\pi/2)$ and $(-\pi/2,-\pi/2)$,
and a maximum at $(\th_1,\th_2)=(0,0)$ \cite{Shima}.
The energy difference between the minimum and the maximum
is determined by the interaction energy
$E_D$ or, equivalently, the parameter $\Gamma=E_D/E_K$.
When $\Gamma$ is much larger than unity,
the energy difference becomes so large that the initial eigenstate
is strongly localized at the two potential minima,
as shown in Fig.~4 (i).
In addition, the amplitude of the wavefunction at the origin becomes almost zero
due to the large potential maximum.
For post-kicked time, however, the kick creates
a large number of excited states,
so that a superposition of them can produce
a transient angular focusing at the potential maximum $(\th_1,\th_2)=(0,0)$.
This leads to a minimal orientation factor at a focal time.
On the other hand, in the {\it classical} limit,
the orientation of paired rotors in the field direction $(\th_1,\th_2)=(0,0)$
cannot occur when the energy difference between the minimum and the maximum
is larger than the kinetic energy of the rotors
immediately after the kick.
In other words, the strong dipolar interaction prevents the paired-rotor state
from being located at the origin $(\th_1,\th_2)=(0,0)$.
This follows that 
the enhanced orientation of paired rotors is a purely quantum phenomenon.

%----------------------------------------------------------------------------------
\begin{figure}[ttt]
\includegraphics[width=8.2cm]{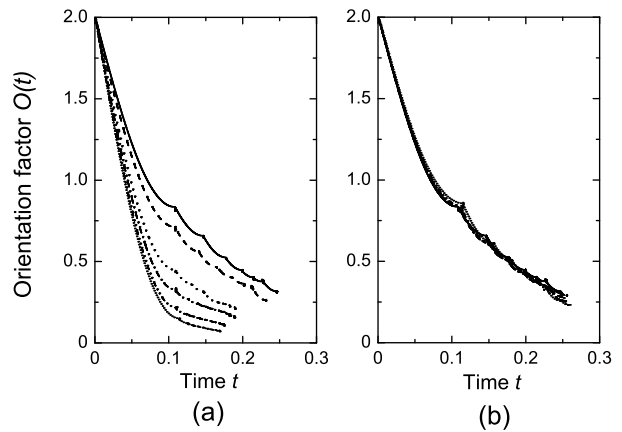}
\caption{ The orientation factor $O(t)$ for paired rotors
kicked with a sequence of seven pulses of the strength $P=10$;
The values of $\Gamma$ are varied as 
$\Gamma=0,\; 1.0,\; 3.0,\; 5.0,\; 10.0,\; 30.0$ from the top (the solid line)
to the bottom (the dotted one).
The quantity $\hbar/E_K$ is taken as unit of time.
}
\label{fig5}
\end{figure}
%--------------------

\subsection{The accumulative squeezing}

The orientation of paired rotors
can be further enhanced
by applying the "accumulative squeezing" scheme proposed in Ref.\cite{Averbukh_2001}.
This strategy is based on a specially designed series of short laser pulses
leading to a dramatic narrowing of the rotor angular distribution.
Figure 5 shows the orientation factor $O(t)$ of paired rotors
kicked by a sequence of seven pulses of the strength $P=10$.
The values of $\Gamma$ increase from
$0$ (solid) to $30.0$ (dotted).
For both arrangement, the strategy works well
to achieve the angular squeezing in paired rotors.
Moreover, in arrangement (a),
a considerable reduction of the factor $O(t)$ is seen for 
strongly interacting paired rotors with $\Gamma \gg 1$.
This result provides 
a new prospect for the scheme of multiple-pulse angular squeezing
in interacting quantum rotors.

% =========================================

\section{Conclusion}

In conclusion, we have theoretically investigated the
quantum dynamics of paired kicked rotors.
The orientation of paired rotors
after the $\delta$-function kick is remarkably enhanced by introducing
dipolar interaction between rotors, 
when the rotors are deposited in an identical plane.
The enhanced orientation is attributable mainly to two factors:
i) The initial state before the kick consists of two wave packets
strongly confined at symmetric positions with respect to the origin 
in the $\th_1$-$\th_2$ space;
ii) These wave packets move translationally toward the origin after the kick.
We have also demonstrated that the orientation of quantum paired rotors 
can be further enhanced by applying a specially designed sequence of pulses.
Our findings will stimulate experimental works
aimed at the orientation of polar molecules correlated via dipolar interaction.

% =========================================

\begin{acknowledgments}

We acknowledge the support for this research by a Grant--in--Aid for Scientific
Research from the Ministry of Education, Science, Sports and Culture, Japan.
Numerical calculations were performed on the Hitachi SR8000 machine
at the Supercomputer Center, ISSP, University of Tokyo.

\end{acknowledgments}

% =========================================

\appendix

\section{Expansion of Mathieu functions}

Four kinds of Mathieu functions,
${\rm ce}_{2n}$, ${\rm se}_{2n+1}$, ${\rm ce}_{2n+1}$, and ${\rm se}_{2n+2}$,
can be expressed in terms of the Fourier expansion as follows \cite{Whittaker}:
\begin{eqnarray}
{\rm ce}_{2n}(\alpha,v_{\alpha}) 
&=& \sum_{m=0}^{\infty} A_{2m}^{(2n)}(v_{\alpha})\cos 2m \alpha.
\label{eqa01} \\
{\rm se}_{2n+1}(\alpha,v_{\alpha}) 
&=& \sum_{m=0}^{\infty} B_{2m+1}^{(2n+1)}(v_{\alpha})\sin (2m+1)\alpha.
\label{eqa02} \\
{\rm ce}_{2n+1}(\alpha,v_{\alpha}) 
&=& \sum_{m=0}^{\infty} A_{2m+1}^{(2n+1)}(v_{\alpha})\cos (2m+1)\alpha.
\label{eqa03} \\
{\rm se}_{2n+2}(\alpha,v_{\alpha}) 
&=& \sum_{m=0}^{\infty} A_{2m+2}^{(2n+2)}(v_{\alpha})\sin (2m+2)\alpha.
\label{eqa04}
\end{eqnarray}
By substituting Eqs.~(\ref{eqa01}-\ref{eqa04}) into Eq.~(\ref{eq09}),
we obtain successive relations that determine the expansion coefficients.
For $\left\{A_{2m}^{(2n)}\right\}$, as an example, we obtain the following relation:
\begin{eqnarray}
& &\ep A_{0}^{(2n)} - v A_{2}^{(2n)} = 0, \label{eqa05} \\
& &(\ep-4) A_{2}^{(2n)} - v \left( 2 A_0^{(2n)}-A_4^{(2n)} \right) = 0, \label{eqa06} \\
& &(\ep-4m^2) A_{2m}^{(2n)} - v \left( A_{2m-2}^{(2n)}-A_{2m+2}^{(2n)} \right) = 0. 
\quad (m\ge 2) \label{eqa07}
\end{eqnarray}
The orthogonality of the Mathieu functions is described by
\begin{eqnarray}
\int_{-\pi}^{\pi} d\alpha \;\; \mbox{ce}_l(v_{\alpha},\alpha) 
\mbox{ce}_{l'}(v_{\alpha},\alpha) &=& \pi\delta_{ll'}, 
\label{eqa08} \\
\int_{-\pi}^{\pi} d\alpha \;\; \mbox{se}_l(v_{\alpha},\alpha) 
\mbox{se}_{l'}(v_{\alpha},\alpha) &=& \pi\delta_{ll'}, 
\label{eqa09} \\
\int_{-\pi}^{\pi} d\alpha \;\; \mbox{ce}_l(v_{\alpha},\alpha) 
\mbox{se}_{l'}(v_{\alpha},\alpha) &=& 0.
\label{eqa10}
\end{eqnarray}

\section{Explicit form of Eq.(\ref{eq13})}

The explicit form of Eq.~(\ref{eq13}) can be obtained by expanding the state
$\Ps(\xi,\eta,\tau^{+})$ in terms of the Mathieu functions.
Using these relations, the function (\ref{eq13}) is expanded as
\begin{equation}
\Ps(\xi,\eta,\tau^{+}) = \sum_{l=0}^{\infty} \sum_{l'=0}^{\infty}
D_{ll'} f_l (\xi,v_{\xi}) g_{l'}(\eta,v_{\eta}),
\label{eqb01}
\end{equation}
where each type of the Mathieu function is abbreviated as
\begin{eqnarray}
f_{2l} (\xi,v_{\xi}) &=& {\rm ce}_l (\xi,v_{\xi}), \label{eqb02}\\
f_{2l+1} (\xi,v_{\xi}) &=& {\rm se}_l (\xi,v_{\xi}); \quad l=0,1,2 \cdots.
\label{eqb03}
\end{eqnarray}
The definition of $g_{l'}(\eta,v_{\eta})$ is the same as that of $f_{l}$.
The expansion coefficients $\{D_{ll'} \}$ are calculated straightforwardly as
\begin{equation}
D_{ll'}= \int_{-\pi}^{\pi} d\xi \int_{-\pi}^{\pi} d\eta \;\;
\Ps(\xi,\eta,\tau^{+}) f_l (\xi,v_{\xi}) g_{l'}(\eta,v_{\eta}).
\label{eqb04}
\end{equation}
Substituting Eq.~(\ref{eqb01}) into Eq.~(\ref{eq13}),
we obtain the explicit form of $\Ps(\xi,\eta,\tau^{+}+t)$ as
\begin{eqnarray}
\Ps(\xi,\eta,\tau^{+}+t) &=& \sum_{l=0}^{\infty} \sum_{l'=0}^{\infty}
\exp\left[ -\frac{i}{\hbar} (E_{\xi}^{(l)}+E_{\eta}^{(l')})t \right] \nonumber \\
&\times& D_{ll'} f_l (\xi,v_{\xi}) g_{l'}(\eta,v_{\eta}).
\label{eqb05}
\end{eqnarray}
In actual calculations for Eq.~(\ref{eqb05}),
the double summation with respect to $l$ and $l'$
can be truncated at a finite value,
because the expansion coefficient $D_{ll'}$
rapidly decay with increasing $l$ and $l'$.

%%%%%%%%%%%%%%%%%%%%%%%%%%%%%%%%%%%%%%%%%%%%
%
%     References
%
%%%%%%%%%%%%%%%%%%%%%%%%%%%%%%%%%%%%%%%%%%%%


\begin{thebibliography}{99}
\bibitem{experiment}
Surface-mounted polar molecules having an in-plane dipolar moment
and dipole rotors attached to a one-dimensional solid, respectively,
are thought to be candidates for realizing arrangements (a) and (b) in Fig.~1.
\bibitem{initial}
For arrangement (a), 
the initial state of the rotors just before the $\delta$-kick
is set to $\th_1(0)=\th_2(0)=\pi/2$ or $-\pi/2$ 
with the same probability.
For case (b), $\th_1(0)$ can take any value with a uniform probability,
while $\th_2(0)$ is determined by the relation $\th_2(0)=\pi-\th_1(0)$.
The initial angular velocity of $\th_i$ immediately after the kick is determined
by $\dot{\th}_i (0^+) = -(P/I)\sin \th_i(0) $ in both (a) and (b).
\end{thebibliography}
\end{document}